\documentstyle[12pt]{article}
 \advance\oddsidemargin by -1in
 \advance\evensidemargin
 by -1in
 \marginparsep
 0.4cm
 \advance\topmargin by
 -0.0in
 \makeindex

 \newcommand\beq{\begin{equation}}
 
 \newcommand\eeq{\end{equation}}
 \newcommand\beqn{\begin{eqnarray}}
 \newcommand\eeqn{\end{eqnarray}}
 \newcommand{\doublespace} {
 \renewcommand{\baselinestretch} {1.6}
 \large\normalsize}

\begin{document}
\vspace*{3cm}

\begin{center}
{\Large{\bf SOFT COMPONENT OF HARD REACTIONS
\medskip
\newline AND NUCLEAR SHADOWING
\medskip
\newline (DIS, Drell-Yan reaction, heavy quark production)}
\footnote{Invited talk at the Workshop Hirschegg'95:
Dynamical Properties of Hadrons in Nuclear Matter.\newline
In proceedings, ed. by H.~Feldmeier and W.~N\"orenberg,
Darmstadt, 1995, p. 102}}\\
 
\bigskip
 
{\Large Boris Kopeliovich} \\
 \date{}
 
\bigskip
 
 {\it Max-Planck Institut f\"ur Kernphysik,\\ Postfach 103980,
69029 Heidelberg, Germany}\\
 
 and\\
 
 {\it Joint Institute for Nuclear
Research\\ Dubna, 141980 Moscow Region, Russia}\\
 
\vspace{3cm}
 
{\bf Abstract}
\bigskip

\parbox{13cm}
{ \baselineskip 14pt
We consider deep-inelastic lepton 
scattering, Drell-Yan lepton pair and
heavy quark production in the reference frame of the target.
These reactions traditionally treated as hard
 have,
however, a substantial 
leading twist soft contribution.
One of the manifestations of such a
soft component,
 nuclear shadowing, is overviewed.}
\end{center}
 
\newpage
\doublespace

 \normalsize
\section{Introduction}

A parton-model interpretation of a high-energy reaction
depends upon a
reference frame.  For instance, the shadowing in the
 nuclear structure
function measured in deep-inelastic
 scattering (DIS) at small $x$ looks
in the infinite
momentum
 frame of the nucleus like a fusion
 of the parton clouds of the bound nucleons 
 \cite{kancheli}.  On the other hand, one can
view this shadowing as inelastic corrections \cite{gribov} in the
 rest frame
of the nucleus (see for example \cite{book}).  To be
 convinced that
 they are the same, one can
compare these mechanisms of shadowing
 either with the Feynman graphs, or with the
Reggeon diagrams.  Only observables may be
Lorentz invariant,
 but not a space-time interpretation.
 
\medskip
 
 Another example is the Drell-Yan (DY) reaction of lepton pair
production \cite{dy}.  The very structure of the DY formula led
 to a wide
spread opinion that this mechanism corresponds to
 instantaneous lepton pair
production in quark-antiquark
 annihilation independently of a reference
frame. According to the
 factorization theorem the cross section of $l\bar l$
pair
 production is proportional to the subprocess cross section,
 $q\bar q
\rightarrow l\bar l$, times the hadronic distribution
 functions which are the
probabilities to find a quark and an
 antiquark with definite momenta and
virtualities in the colliding hadrons.  
From this point of view the nuclear suppression of
 lepton pair production can be treated as a shadowing in the nuclear
 structure function.
 
 However, even the very statement that the annihilating quark and
antiquark belong to the beam and to the target
 respectively (or vise versa)
is not Lorentz invariant. For instance, in
the target rest frame both
 $q$ and $\bar q$, as well as the lepton pair,
should be
 considered as a beam hadron fluctuation, related to the beam
distribution function, and vise versa in the beam rest frame.
 Correspondingly,
the DY process can be treated as an electromagnetic
 bremsstrahlung of heavy photons by the beam
or the target, depending on a
 reference frame.  Different partonic interpretations of
DY reaction in
 different reference frames correspond,
nevertheless, to the same Feynman
graphs.
 
\medskip
 
 Heavy quark production is also treated as a hard reaction,
subject to perturbative QCD \cite{nason}.  However, similar to
 the DY case,
one is free to choose any reference frame
depending on convenience. Correspondingly, the partonic
interpretation
varies: the subprocess looks like $q\bar q$
annihilation or gluon fusion in the rest frame of the $Q\bar Q$
 pair,
however, it should be treated as a heavy gluon bremsstrahlung,
 looking from the
infinite momentum frame of the $Q\bar Q$.
 
\medskip
 
 In this paper we consider all these processes in the rest frame
of the target and use the light-cone representation for the
 projectile wave
function, which allows to take explicitly into
 account the color-coherence
effects, missed in the standard
 perturbative consideration. We come to the
conclusion that these
 processes have much in common, and all of them have a
substantial
 soft component, which scales in $Q^2$ and 
manifests itself particularly in nuclear
shadowing. Bjorken and Kogut were first who claimed existence
 of such a
component in DIS and its importance for the $Q^2$
 scaling \cite{bk}. A
hadronic fluctuation of the virtual
 photon have to have a low intrinsic
transverse momentum to
 interact softly. Such a fluctuation
 produces jets
aligned along the photon momentum.
 This idea was discussed in the framework
of QCD in \cite{fs1} and
 got its explicit formulation in the 
line-cone
 wave functions formalism 
\cite{nz91}. Analogous small-$k_T$ component of
projectile fluctuations was found in DY reaction and heavy quark
 production
in \cite{bhmt}, and a light-cone formalism was
 developed in \cite{bkz,hk}.

 We overview briefly in what follows the light-cone wave function
 formalism
for the above hard reactions, emphasizing
 manifestations of the soft
component.
 
\section{Scaling variable for nuclear shadowing}

 In spite of the fact that DIS probes only the quark distribution
 function,
there is wide spread believe that shadowing in the
 qluon distribution
function is approximately the same as observed
 in DIS.
 This
 could be
true, if electromagnetic probe of gluon distribution
 were really hard.
However this is not the case.
 Indeed, a virtual photon participates in strong
interaction
 through its hadronic fluctuations.
 Assuming that the transverse
separation $\rho$ of the
 $q\bar q$ fluctuation of a highly virtual photon is
small,
 one can use the property of color
 transparency of the dipole cross
section \cite{zkl} $\sigma(\rho)
 \propto \rho^2$.  If this is the case, the
total photoabsorption
 cross section \cite{nz91} is related to the gluon
distribution
 function $xG_N(x,Q^2)=g_N(x,Q^2)$ \cite{barone}
 
 \beq
\sigma^{\gamma^*N}_{tot}(x,Q^2)=
 \int_0^1d\alpha\int d^2\rho|\Psi_{q\bar q}
(\rho,\alpha)|^2\ \sigma(\rho,x)\approx
 \frac{\pi^2}
 {3}\alpha_s(\rho)\
\langle\rho^2\rangle\ g_N(x,Q^2)
\label{1}
\eeq
 
 Here $\alpha$ is the relative fraction
 of the photon light-cone
momentum carried by the quark.
 Hereafter we neglect the small contribution of
the longitudinal
 component of the photon wave function and the quark mass
unless
 it is important.  The wave function reads \cite{nz91},
 
 \beq
|\Psi_{q\bar q}(\rho,\alpha)|_T^2\approx
\frac{6\alpha_{em}}{(2\pi)^2}\sum_1^{N_f}Z_i^2
 [1-2\alpha(1-\alpha)]\
\epsilon^2K_1^2(\epsilon\rho)\ ,
\label{2}
\eeq
 where $K_1(z)$ is modified Bessel function,
 
 \beq
\epsilon^2=\alpha(1-\alpha)Q^2+m_q^2
\label{3}
\eeq
 
 According to (\ref{2}), (\ref{3}) the mean $q-\bar q$ transverse
separation squared, $\rho^2\propto 1/\epsilon^2$, is of the order
 of $1/Q^2$,
except the edges of the kinematical region $\alpha$ or
 $1-\alpha\sim
m_q^2/Q^2$, where the $q\bar q$ fluctuation
 acquires a large transverse
separation, $\rho^2\sim 1/m_q^2$ (actually, the quark mass should
be replaced by an effective cut off of the order of $\Lambda_{QCD}$).
 The presence of this soft component in the
$q\bar q$ fluctuation
 of the photon, makes questionable the validity of the 
electromagnetic probe for the gluon distribution.  Indeed, if one
 needs, for
example, to figure out how many nucleons are in a
 nucleus, and one uses, say,
a pion-nucleus inelastic interaction
 as a probe, one obviously gets a
wrong result, namely, the
number of nucleons $\sim A^{2/3}$.  The source of
the trouble is
 the softness of the probe, $\sigma_{tot}(\pi N)$ is too
large.
 To
 get the correct answer one should use a probe with a
sufficiently
 small cross section, $\sigma\ll 1/R_A\rho_A$, where $R_A$ and
$\rho_A$ are the radius and the density of the nucleus.
 
 The same problem
arises for probing the gluon distribution 
by a photon, due to
asymmetric $q\bar q$
fluctuations with large transverse
 separations, which experience 
shadowing, interacting with a
 cloud of gluons.  Such shadowing looks like 
a suppression of the photoabsorption
 cross section on the nucleus, but it still does not mean 
that the gluon density at small $x$ is reduced in the nucleus. 
 One should
disentangle a nuclear suppression of gluon density at small $x$
 and a
regular, Glauber-like shadowing of the photon fluctuations
propagating through the
nucleus. Both, however, depend on the density of gluons which
 originate from
different nucleons and overlap.
 
 Let us test universality of the relation
between shadowing and the
 gluon density \cite{kp} on a sample of available data
on DIS on
 nuclei.  We start with the general expression for the
photoabsorption cross section on a nucleus \cite{kl78,zkl,nz91},
 which takes
into account all the inelastic shadowing corrections
 \cite{gribov}.
 
 \beq
\sigma^{\gamma^*A}_{tot}(x,Q^2)=
 2\int d^2b\left\langle 1-
 \left[1-
\frac{\sigma(\rho,x)T(b)}{2A}\right]^A
 \right\rangle
\label{4}
\eeq
 
 The partial amplitude of $q\bar q$ elastic scattering is averaged
over the eigenstates of interaction, which have 
a definite transverse sizes.  In the
case of photoabsorption the averaging is weighted
 with the photon wave
function squared (\ref{2}), in analogy to
 (\ref{1}).  Nuclear thickness
$T(b)\approx
 \int_{-\infty}^{\infty}dz\rho_A(b,z)$.
 
 Making use of expansion
in r.h.s of (\ref{4}) one can represent
the nuclear shadowing effects in the
form,
 
 \beq
 R^{DIS}_{A/N}(x,Q^2)=\frac{ \sigma^{\gamma^*A}_{tot}(x,Q^2)}
{A\sigma^{\gamma^*N}_{tot}(x,Q^2)}=
 1-{1\over
4}\frac{\langle\sigma^2(\rho,x)\rangle}
 {\langle\sigma(\rho,x)\rangle}
 \langle T(b)\rangle
 +...\ ,
\label{5}
\eeq
 were $\langle T(b)\rangle=(1-1/A)/A\int d^2bT^2(b)$.
 The cross section of
interaction
 of a $q\bar q$ fluctuation 
 with a nucleon in 
can be represented in the form \cite{bfkl},
$\langle\sigma(\rho,x)\rangle \approx 
\langle\sigma(\rho)\rangle/x^{\Delta_P(Q^2)}$,
where $\Delta_P(Q^2)$ measured at HERA \cite{zeus,h1}
reaches a large value $0.3 \div 0.4$ at high $Q^2$.
On the other hand, $\langle\sigma^2(\rho,x)\rangle$ is 
dominated by the soft asymmetric fluctuations, and its
$x$-dependence is governed by the soft Pomeron,
$\langle\sigma^2(\rho,x)\rangle \approx
\langle\sigma^2(\rho)\rangle/x^{2\Delta_{soft}}$,
where $\Delta_{soft} \approx 0.1$. Correspondingly,
$\langle\sigma^2(\rho)\rangle$ has a leading twist
behaviour $\propto 1/Q^2$.
Therefore, one can write

 \beq
\frac{\langle\sigma^2(\rho,x)\rangle}
 {\langle\sigma(\rho,x)\rangle} =
\frac{N}{F_2^p(x,Q^2)}
\ \left({1\over x}\right)^{2\Delta_{soft}}\ ,
\label{6}
\eeq

\noindent
where N is a free parameter.

 An important ingredient of formula (\ref{4}) is the assumption
 that
$x$ is sufficiently small, $x\ll 1/m_NR_A$, which
guarantees that
all the parton clouds of
nucleons with the same impact
 parameter overlap in the antilaboratory frame. 
This is the same as to say that
 the lifetime of the photon fluctuation is
 much longer than the
nuclear radius.
 However, most of
 available data are in the 
region of $x$, where this condition
 is broken.
Finiteness of the lifetime,
which is usually called coherence time, can be 
taken into account
taking into account the phase shifts
between $q\bar q$ wave packets
 produced at different longitudinal
coordinates, which
 suppress the effective nuclear thickness in
 (\ref{5}),
 
\beq
 \langle\widetilde{T}(b)\rangle={1\over A}\int d^2b
\left[\int_{-\infty}^{\infty}dz\ \rho_A(b,z)\ e^{iqz}\right]^2
\label{7}
\eeq
 
 Here $q=(Q^2+M^2)/2\nu$ is the longitudinal momentum transfer
in the diffractive photoproduction of a hadronic state with mass $M$.
 
Available data on nuclear shadowing are taken at different
values of $x$ and $Q^2$, even within the
 same experiment, what makes it difficult to compare with
 theoretical
calculations.  Since we expect the shadowing effects to
be dependent only on the amount of
overlapping gluons, the data should be
 plotted against a new variable
$n(x,Q^2)$,
 
 \beq
 n(x,Q^2)=
 \frac{N}{4F_2^p(x,Q^2)}
 \langle\widetilde T(b)\rangle
\ \left({1\over x}\right)^{2\Delta_{soft}} \ ,
\label{8}
\eeq
 which has to be calculated for each experimental point.
 
Although one may have a plausible guess about 
the value of parameter $N$,
calculation is quite
 ambiguous.  Instead, we intend to test the scaling
 of nuclear shadowing versus new variable
$n(x,Q^2)$, which is $N$-independent.

 We calculated $n(x,Q^2)$ for
each point of
 data \cite{nmc1,nmc2} for $R^{DIS}_{A/N}(x,Q^2)$ and plotted
them
 against this variable in fig.  1.  The data demonstrate
 excellent
scaling in $n(x,Q^2)$.  The absolute value of shadowing
fixes the parameter $N\approx 3\ GeV^{-2}$.
 
\begin{figure}[tbh]
\includegraphics{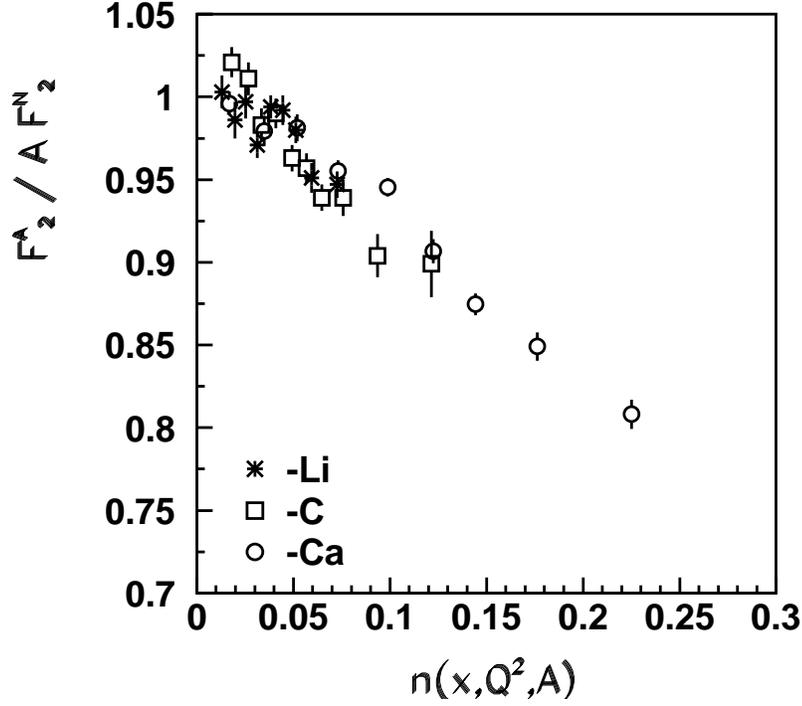}
\begin{center}
\vspace{9cm}
\parbox{13cm}
 {\caption[Delta]
 {Normalized by $1/A$ 
nucleus-to-nucleon ratio of the structure functions
plotted versus scaling variable $n(x,Q^2)$, defined in
 (\ref{8}). The data
are from \cite{nmc1,nmc2}.}
\label{fig1}}
\end{center}
\end{figure}

Notice that the data with $Q^2\ll m_{\rho}^2$ is
subject to vector
 dominance model, they expose the same nuclear shadowing as
real
 photons, independently on $x$. This is the reason of saturation
 of
nuclear shadowing at low $x$, claimed in \cite{e665,nmc2}.
 We excluded the data points \cite{nmc2} with
 $Q^2<0.5\ GeV^2$ from
the analyses.
 
\section{Nuclear shadowing of Drell-Yan lepton pairs}

 The Born diagrams of perturbative QCD describing the quark
(antiquark)-nucleon interaction with radiation of a heavy photon,
 converting
into a lepton pair of mass $M$, are shown in fig. 2.

\begin{figure}[tbh]
\includegraphics{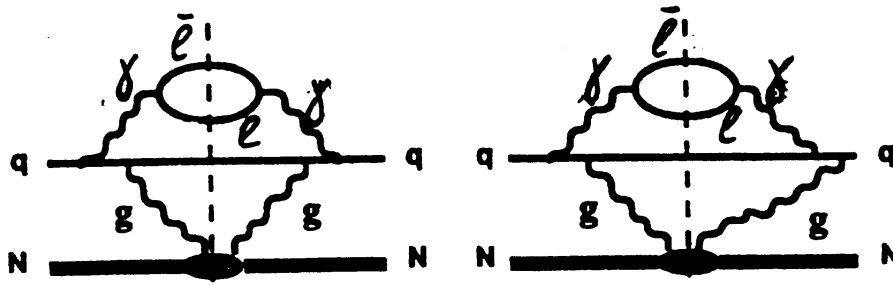}
\begin{center}
\vspace{4cm}
\parbox{13cm}
 {\caption[Delta]
 {Born diagrams for lepton pair
production.}
\label{fig1}}
\end{center}
\end{figure}

 We are interested in
moderate values of $x_1$ and small $x_2\ll
 1$, where $x_{1,2}=\pm
x_F/2+\sqrt{x_F^2/4+M^2/s}$, what
 corresponds to gluonic exchanges.  The
calculation of these
 diagrams in impact parameter representation \cite{bkz}
leads to
 the following expression for the cross section of lepton pair
production in $hN$ interaction,
 
 \beq
M^2\frac{d\sigma_{DY}^{hN}}{dM^2dx_1}=
 \int_{x_1}^1d\alpha
F_q^h(\frac{x_1}{\alpha})
 \int d^2\rho\;
 \left|\Psi_{ql\bar
l}(\alpha,\rho)\right|^2\;
 \sigma_{l\bar l}(q\rightarrow ql\bar l)\ .
\label{2.1}
\eeq
 
 Here $F_q^h(z)$ is the quark distribution function of the hadron.

 \beq
 \left|\Psi_{ql\bar l}
(\alpha,\rho)\right|^2\approx\frac{(Z_q\alpha_{em})^2}{\pi^2}\
 Im\
\Pi(M^2)\;\alpha\tau^2K_1^2(\tau\rho)\ ,
\label{2.2}
\eeq
 is the light-cone wave function squared of the
 $ql\bar l$ Fock
component of the projectile quark in the mixed
 $\rho-\alpha$ representation,
where $\rho$ is the transverse
 separation between $q$ and the center of
gravity of the $l\bar
 l$-pair, and $\alpha$ the fraction of the projectile
quark
 light-cone momentum carried by the $l\bar l$ pair.
The value of $\Pi(M^2) \approx 1/12\pi$ was calculated in
\cite{bhmt}.

 There is a close
similarity between wave function (\ref{2.2})
 and that for the $q\bar q$
fluctuation of a photon (\ref{2}). The
 parameter $\tau$ is also close to
$\epsilon$,
 defined in (\ref{3}),
 $\tau^2=(1-\alpha)M^2+\alpha^2m_q^2$
 
Most surprisingly, the cross section of freeing of the $l\bar
 l$-pair turns
out to be the same dipole interaction cross
 section of a colorless $q\bar
q$ pair with a separation
 $\alpha\rho$, $ \sigma_{l\bar l}(q\rightarrow
ql\bar l)=
 \sigma(\alpha\rho,x_2)$.
  This has a natural interpretation
\cite{bkz}.  The color
 screening factor $[1-\exp(i\vec q\vec r)]$ in the
dipole
 interaction cross section $\sigma(\rho)$ 
\cite{zkl} originates from the
 difference
in the impact parameters of gluon attachments to $q$
 or $\bar q$.  The same
is valid for the graphs in fig. 2.  The
 quark trajectories before and after
the photon radiation have
 different impact parameters.  The $q\gamma*$
fluctuation with
 transverse separation $\rho$ has a center of gravity,
which
 coincides with the impact parameter of the parent quark.  It is
separated by $(1-\alpha)\vec\rho$ from the $\gamma^*$ and
 $\alpha\vec\rho$
from the $q$. Thus, the diagrams in fig.  2
 have relative phase factor
$\exp(i\alpha\vec\rho\vec q)$ and
 different signs, what leads to the dipole
form of the cross
 section.
 
\medskip
 
 The cross section of DY pair production off a nuclear target can
be written in analogy to (\ref{4}),
 
 \beq
M^2\frac{d\sigma_{DY}^{hA}}{dM^2dx_1}=
 2\int d^2b\left\langle 1- \left[1-
\sigma(q\rightarrow
 ql\bar l)\frac{ T(b)}{2A}\right]^A \right\rangle\ ,
\label{2.4}
\eeq
 where averaging over $\alpha$ and $\rho$ is weighted in the same
 way
as in (\ref{2.1}).
 
 The corrections for finiteness of the lifetime of
the
 $ql\bar l$ fluctuation can be done in the same way as in the case
 of
DIS. The coherence time turns out to be about the same
 \cite{bkz},
$l_c=1/q\approx 1/2x_2m_N$.

Keeping in (\ref{2.4}) the first-order shadowing correction
we find for the nucleus-to-nucleon ratio of DY cross sections

\beq
R^{DY}_{A/N}(x_F,M^2) \approx
1 - n[x_2,M^2(1-x_1)]\ ,
\label{2.5}
\eeq
where $n(x,Q^2)$ is defined in (\ref{8}). 

Comparing (\ref{2.5}) with (\ref{5}) we see that
the factorization theorem should be modified: 
DIS and DY structure function are to be
 compared at $Q^2=M^2(1-x_1)$. The reason for this correction
is clear: the average $q\gamma^*$ 
fluctuation becomes more and more asymmetric
at $x_1 \to 1$, since integration over $\alpha$
in (\ref{2.1}) goes from $x_1$ to $1$.
Thus, the projectile fluctuations in DY reaction are
more asymmetric, i.e. are softer, than in DIS.
This provides a deviation from factorization and 
should be taken into account.
 
\begin{figure}[tbh]
\includegraphics{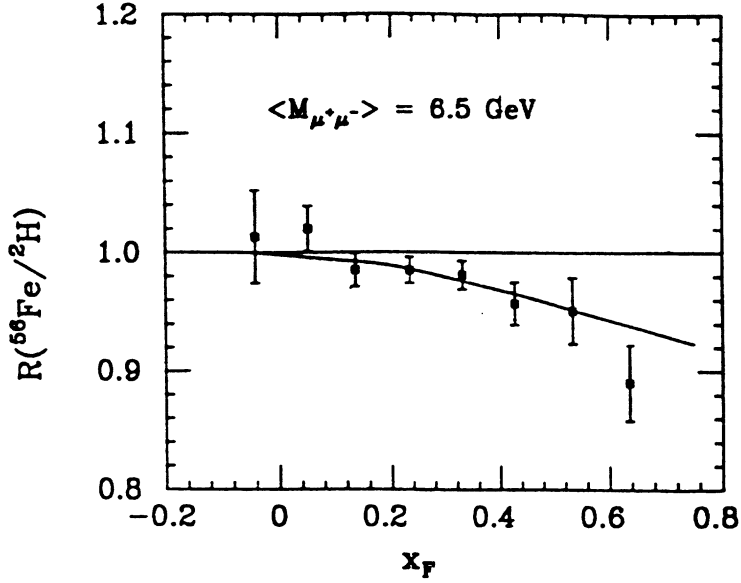}
\begin{center}
\vspace{7.2cm}
\parbox{13cm}
 {\caption[Delta]
 {Comparison of the parameter-free calculation of
nuclear shadowing in DY reaction with the E772 data.}
\label{fig1}}
\end{center}
\end{figure}

As soon as parameter $N$ is fixed by the data on nuclear
shadowing in DIS, one can predict the nuclear suppression 
of DY lepton pairs.  Comparison with the
results of the E772 collaboration \cite{e772} at
 $800~GeV$ shown in fig. 3
demonstrates a good agreement.
 
Note that the increase of nuclear shadowing towards
$x_F=1$ is mainly due to the increase of the fluctuation
lifetime $t_c$, similar to DIS.

\section{Heavy quark production}

 Analogous to DY reaction the standard approach to heavy quark
 production
is based on the factorization theorem and uses the
 quark and gluon
distribution functions of the colliding hadrons.
  Heavy quark pair production may also be treated, 
similar to the DY lepton
 pairs, as a
bremsstrahlung in the infinite momentum frame.  In
 this case we have the same diagrams as in fig. 2, with
replacement of $\gamma^*\rightarrow l\bar l$ by $g^*\rightarrow
 Q\bar Q$, and
on top of that a few new ones shown in fig.  4.
 These new Feynman graphs correspond to direct
interaction of the virtual gluon
 and heavy quarks with the target.  Comparing
with the standard
 approach one may say that those diagrams on fig.  2
correspond to
 $q\bar q$ annihilation, while ones in fig.  4 correspond to
gluon
 fusion mechanism.  The vertex $gg\rightarrow Q\bar Q$ in fig.  4a
 is
decomposed explicitly in fig.  4b.
 
\begin{figure}[tbh]
\includegraphics{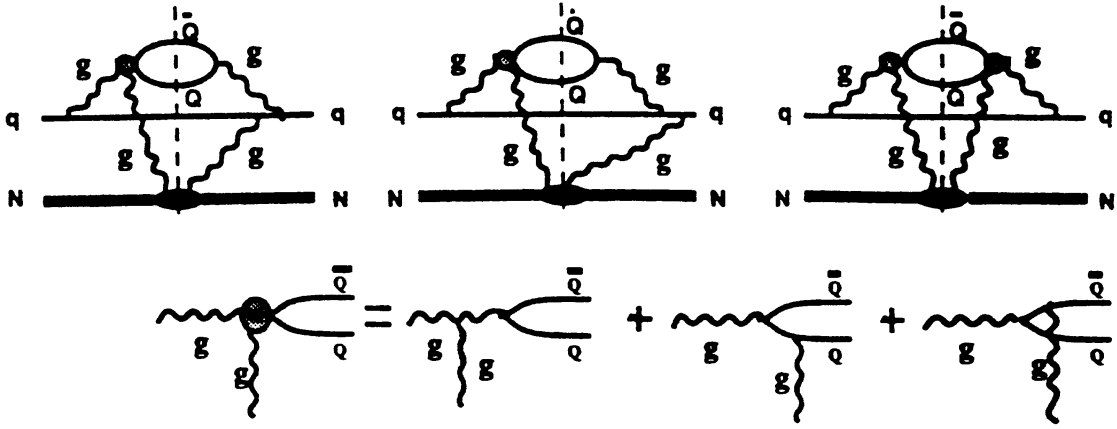}
\begin{center}
\vspace{5cm}
\parbox{13cm}
 {\caption[Delta]
 {The additional to Fig. 2
Born diagrams for heavy quark production.}
\label{fig1}}
\end{center}
\end{figure}

 The important result of the previous
section, the cross section of
 lepton pair production in form (\ref{2.1}), can
be naturally
 extended to a general case of a virtual decay of a parton
$a\rightarrow b+c$, where $a, b, c$ may be colored or colorless
 \cite{hk}.
The cross section of freeing the fluctuation $bc$, which
has a transverse 
 separation $\vec\rho$
and relative fraction $\alpha$ of the longitudinal
 momentum carried by $b$, equals to the
dipole total cross section
 of interaction of a colorless system $\bar abc$
with relative
 transverse separations, $\vec\rho$ $(bc)$,
$(1-\alpha)\vec\rho$
 $(b\bar a)$ and $\alpha\vec\rho$ $(\bar ac)$:
$\sigma(a\rightarrow bc)=\sigma_{\bar
 abc}[\rho,(1-\alpha)\rho,\alpha\rho]$

 The color screening provides the infra-red stability of the
 freeing cross
section, even if $a$ is colored. The cases of
 $\gamma^*\rightarrow q\bar q$
and $q\rightarrow q\gamma^*$ were
 discussed above.  Below we encounter new
examples, like
 $q\rightarrow qg^*$ and $g\rightarrow Q\bar Q$ fluctuations.

 This mnemonic rule applied to the whole set of diagrams in fig.
 3, leads
to the $Q\bar Q$ production cross section, which has a
 structure similar to
that for DY reaction (\ref{2.1}),

 \beqn
 & &M^2\frac{d\sigma_{Q\bar
Q}^{hN}}{dM^2dx_1}=
 \int_{x_1}^1 d\alpha F_q^h({x_1\over\alpha})
 \int
d^2\rho\int d^2r
 \left|\Psi_{qg^*}(\alpha,\rho)\right|^2\times
 \nonumber
\\
 & &\int_0^{\beta_{max}}\;d\beta
 \left|\Psi_{Q\bar Q}(\beta,r)\right|^2
\left[\sigma(q\rightarrow qg)+
 \sigma(g\rightarrow Q\bar Q)\right]\ ,
\label{3.7}
\eeqn
 
 \beqn
 & &\sigma(q\rightarrow qg)=\sigma_{gq\bar
q}[\rho,(1-\alpha)\rho,\alpha\rho]\;, \nonumber \\
 & &\sigma(q\rightarrow
Q\bar Q)=
 \sigma_{gq\bar q}[\beta r,(1-\beta)r,r]\;.
\label{3.8}
\eeqn
 
 Here, $r$ is the transverse separation of the $Q\bar Q$ pair;
$\beta$ is a share of the gluon longitudinal momentum carried by
 $Q$.  The
total cross section for the interaction of a colorless
 $gq\bar q$ system with
a nucleon can be expressed in terms of
 $q\bar q$ dipole cross sections
\cite{nz1}, $\sigma_{gq\bar
 q}(\rho_1,\rho_2,r)= 9/8\left\{\sigma_{q\bar
q}(\rho_1)+
 \sigma_{q\bar q}[(1-y)\rho_2]\right\}-1/8\sigma(r)$.
 
 The
$qg^*$ Fock state wave function squared,
$\left|\Psi_{qg^*}(\tau,\rho)\right|^2$, has the same form
 as
 eq.
(\ref{2.2}), except for a replacement in the normalization,
$(Z_q\alpha_{em})^2\rightarrow(\alpha_s/2N_c)^2$.  We denote by
 $g^*$ a
point-like $Q\bar Q$ pair in the color-octet point-like
 state, which is
indistinguishable from a gluon.  The interaction can resolve
 such a pair only if it has
a separation, i.e.  a color dipole
 moment.
 
 The squared wave function
$\left|\Psi_{Q\bar
 Q}(\beta,r)\right|^2$ of the $Q\bar Q$ fluctuation of a
gluon is
 equal to the photon's one (\ref{2}) up to a simple change of the
normalization.
 
 Interaction with the spectator light quark
 has a
substantial soft component at $1-\alpha\sim m_q^2/M^2$.
 This gives a
contribution to (\ref{3.7}) which is $\propto
F_h^q(x_1)/M^2)\ln[(1-x_1)M^2/m_q^2]$.
 
 The contribution of the direct
interaction was considered also in
 \cite{npz}.  It is suppressed by a factor
$1/M_{Q\bar Q}^2$,
 since $\langle r^2\rangle\sim 1/m_Q^2$.  This contribution
to
 cross section (\ref{3.7}) is $\propto (1-x_1)F_h^q(x_1)/M^2$ and
 is
small compared with the spectator quark interaction.
 
\medskip
 
 The soft interaction component under discussion provides a
substantial nuclear suppression, similar to that in DY reaction.
 If the
lifetime of the $qQ\bar Q$ fluctuation is long in
 comparison with the
nuclear radius, the nuclear cross section in
 frozen approximation reads
similar to (\ref{2.4})
 
 \beq
 M^2\frac{d\sigma_{Q\bar
Q}^{hA}}{dM^2dx_1}=
 2\int d^2b\left\langle 1- \left\{1-
\left[\sigma(q\rightarrow qg)+
 \sigma(g\rightarrow Q\bar Q)\right]
 \frac{
T(b)}{2A}\right\}^A \right\rangle\ ,
\label{3.11}
\eeq
 
 The averaging corresponds to the integration over $\rho$, $r$,
$\alpha$ and $\beta$ in (\ref{3.7}).
 Analogous to DIS and DY reaction this
nuclear
 shadowing of heavy quark production scales with $M^2$.
 
 Typical
value of the predicted effect for charm
 production is about $20-30\%$ for
heavy nuclei. We avoid here
 comparison with available data on open charm
hadron production,
 since it needs to include into consideration the
hadronization
 stage. This is expected to rearrange substantially the
produced hadron momenta \cite{hk}.
 
\bigskip
 
 Summarizing, there is a substantial soft 
contribution to DIS, DY pair and heavy quark production,
which is a leading twist effect.  It has
 a common
origin in all reactions, namely, highly asymmetric
 in longitudinal momentum projectile 
hadronic fluctuations 
are subject to color opacity. These
soft components lead to nuclear shadowing,
 diffractive production. It causes a
deviation from the
factorization.  In this
 mini-review we treated all these reactions in
the same approach and found a close similarity between 
them.


\begin{thebibliography}{10}
\bibitem{kancheli} O.V.~Kancheli, Pisma ZHETF, {\bf 18} (1973)
 469
\bibitem{gribov}
 V.N. Gribov, Sov. Phys. JETP {\bf 29} (1969) 483
\bibitem{book} Yu.L.~Dokshitzer, V.A.~Khoze, A.H.~Mueller and
 S.I.~Troyan,
'Basics of Perturbative QCD', Editions Fronti\'eres,
 1991
\bibitem{dy} S.D.~Drell and T.M.~Yan, Phys. Rev. Lett. {\bf 25}
 (1970) 316
\bibitem{nason} P.~Nason, S ~Dawson and R.K.~Ellis, Nucl. Phys.
 {\bf B303}
(1988) 607; {\bf B327} (1989) 49
\bibitem{bk} J.D.~Bjorken and J.~Kogut, Phys. Rev. {\bf D8}
 (1973) 1341
\bibitem{fs1} L.L.~Frankfurt and M.I.~Strikman, Phys. Rept.
 {\bf 160} (1988)
235
\bibitem{nz91}
 N.N. Nikolaev and B.G. Zakharov, Z. Phys. {\bf C49} (1991)
607
\bibitem{bhmt} S.J.~Brodsky et al., Nucl.  Phys.  {\bf B 369}
 (1992) 519
\bibitem{bkz}
 O.~Benhar, B.Z.~Kopeliovich and A.~Zieminski, 'Soft component
of
 Drell-Yan mechanism of lepton pair production', paper in
 preparation.
\bibitem{hk} J.~H\"ufner and B.Z.~Kopeliovich, 'Color coherence
 in
hadroproduction of heavy quarks from nucleons and nuclei',
 paper in
preparation.
\bibitem{zkl} Al.B.  Zamolodchikov, B.Z.  Kopeliovich and L.I.
 Lapidus, JETP
Lett.  {\bf 33} (1981) 612.
\bibitem{barone} V.~Barone et al., Z. Phys. {\bf C58} (1993) 541
\bibitem{fs} B.~Bl\"attel et al., Phys. Rev. Lett. {\bf
 71} (1993) 896
\bibitem{kp} B.Z.~Kopeliovich and B.~Povh, 'What can we learn
 from nuclear
shadowing about the proton structure function at
 small x?', paper in
preparation
\bibitem{kl78} B.Z.~Kopeliovich and L.I.~Lapidus, Sov. Phys. JETP
 Lett. {\bf
} (1978)
\bibitem{bfkl} E.A.~Kuraev, L.N.~Lipatov and V.S.~Fadin,
 Sov. Phys. JETP {\bf
44} (1976) 443; {\bf 45} (1977) 199;
 Ya.Ya.~Balitskii and L.I.~Lipatov,
Sov. J. Nucl. Phys. {\bf 28}
 (1978) 822; L.N.~Lipatov, Sov. Phys. JETP {\bf
63} (1986) 904
\bibitem{nmc1} CERN NMC, P.~Amaudruz et al., Z.
 Phys.  {\bf C51} (1991) 387
\bibitem{nmc2} CERN NMC, M.~Arneodo et al.,
 submitted to Nucl. Phys.
\bibitem{e665} FNAL E665, M.R.~Adams et al., Phys. Lett.
 {\bf B287} (1992)
375
\bibitem{kz} B.Z.~Kopeliovich and B.G.~Zakharov, Phys. Rev. {\bf
 D44} (1991)
3466
\bibitem{zeus} DESY ZEUS, M.~Derrick et al., DESY 94-143, August
 1994
\bibitem{h1} DESY H1, G.~R\"adel, H1-10/94-390, 1994
\bibitem{e772}
 D. M. Alde {\em et al}, Phys. Rev. Lett. {\bf 64} (1990)
2479.
\bibitem{nz1} N.N.~Nikolaev and B.G.~Zakharov, J\"ulich preprint,
KFA-IKP(Th)-1993-17
\bibitem{npz} N.N.~Nikolaev, G.~Piller and B.G.~Zakharov,
 J\"ulich preprint,
KFA-IKP(TH)-1994-43
 
\end{thebibliography}
\end{document}